\documentclass[twocolumn,superscriptaddress,floatfix,preprintnumbers,prl ,nofootinbib,hyperref,longbibliography]{revtex4-1}
\usepackage{enumerate}
\usepackage{mathrsfs,amsmath,amssymb}
\usepackage{natbib}
\usepackage{graphicx}
\usepackage{cancel}
\usepackage[normalem]{ulem}
\usepackage[colorlinks, linkcolor=red, anchorcolor=blue, citecolor=blue, colorlinks=true, urlcolor=blue]{hyperref}
\hypersetup{colorlinks=true}
\usepackage{stackrel}
\usepackage{pgfplots}
\usepackage{subfigure}
\usepackage{makecell}
\usetikzlibrary{positioning}
\usetikzlibrary{snakes}
\usepackage{multirow}
\usepackage{slashed}
\usepackage{float}
\newcommand{\prlsection}[2]{{\it\textbf{#1}{#2}}---}

\newcommand\be{\begin{equation}}
\newcommand\ee{\end{equation}}
\newcommand\bea{\begin{eqnarray}}
\newcommand\eea{\end{eqnarray}}

\begin{document}

\title{ Eogenesis via the High-scale Electroweak Symmetry Restoration}
\author{Wei Chao}
\email{chaowei@bnu.edu.cn}
\affiliation{Key Laboratory of Multi-scale Spin Physics, Ministry of Education, Beijing Normal University, Beijing 100875, China}
\affiliation{Center of Advanced Quantum Studies, School of Physics and Astronomy, Beijing Normal University, Beijing, 100875, China }
\vspace{3cm}

\begin{abstract}

In this paper, we propose a novel electron-assisted Baryogenesis scenario that does not require explicit B-L violation, which is essential for the traditional Leptogenesis mechanism. This scenario is based on the assumption of high-scale electroweak symmetry restoration, which implies that the electron Yukawa interaction, crucial for the mechanism, does not reach thermal equilibrium before the electroweak sphaleron process is quenched in the early universe. Primordial charge asymmetries for chiral electrons, which can be generated through various mechanisms such as axion inflation, the evaporation of primordial black holes, or the CP-asymmetric decays of a heavy Higgs doublet, serve as the initial condition for the amplification of the baryon asymmetry through transport equations. Right-handed electron asymmetry is almost irrelevant to the baryon asymmetry due to high-scale electroweak symmetry restoration, leading to both a non-zero baryon asymmetry and the electron asymmetry. We dub this mechanism as the Eogenesis.

\end{abstract}

\maketitle
\prlsection{Introduction}{.}  The Standard Model (SM) of particle physics has been remarkably successful, yet it is recognized as an incomplete theory due to the presence of phenomena that suggest new physics beyond the SM. Among these, the observed baryon asymmetry of the universe (BAU) is a significant indicator of the need for a more fundamental theory. 
The BAU is measured at the time of the Big Bang Nucleosynthesis and the Cosmic Microwave Background with~\cite{Planck:2018vyg}
\bea
\eta = \frac{n_b }{s} = 9 \times 10^{-11} \; .
\eea
where $n_b$ and $s$ are the baryon density and the entropy density, respectively. 
According to Sakharov~\cite{Sakharov:1967dj}, a dynamical theory must satisfy the following three conditions to generate the BAU:(1) the baryon number $(\mathbf{B})$ violation; (2) $\mathbf{C}$ and $\mathbf{CP}$ violations; and (3) a departure from the thermal equilibrium.  
In the SM, the electroweak sphaleron process~\cite{Klinkhamer:1984di} violates the baryon number and the $\mathbf{CP}$ symmetry is violated in the CKM matrix, but the third condition is not satisfied.

Many workable {\it Baryogenesis} mechanisms have been proposed,  of which the Leptogenesis~\cite{Fukugita:1986hr}, Electroweak Baryogenesis (EWBG)~\cite{Cohen:1993nk,Trodden:1998ym,Morrissey:2012db},  Afleck-Dine Baryogenesis~\cite{Affleck:1984fy} and Spontaneous Baryogenesis~\cite{Cohen:1988kt},  are representative great ideas in the intersection of the energy,  the intensity and  the cosmic frontiers.  
Besides, there are  also new ideas about the {\it Baryogenesis}, including EWBG with spontaneous CP violations~\cite{Chao:2017oux}, axiogenesis~\cite{Co:2019wyp,Domcke:2020kcp}, axion-inflation {\it Baryogenesis} (AIB)~\cite{Domcke:2019mnd}, QCD {\it Baryogenesis}~\cite{Croon:2019ugf}, Hylogenesis~\cite{Davoudiasl:2010am}, Darkogenesis~\cite{Shelton:2010ta}, WIMP-Triggered {\it Baryogenesis}~\cite{Cui:2011ab}, freeze-in {\it Baryogenesis}~\cite{Hall:2010jx}, Mesogenesis~\cite{Elor:2018twp}, Majorogenesis~\cite{Chao:2023ojl}, etc.  
For further details, the reader is referred to Ref.~\cite{Elor:2022hpa} and the references cited therein.
%
%
In the past few decades, the Leptogenesis mechanism has received widespread attention due to the flourishing development of the neutrino physics, as it addresses the BAU via the CP-asymmetric decay of heavy seesaw particles, which are applied to explain the tiny active neutrino masses via the so-called seesaw mechanism~\cite{Minkowski:1977sc,Yanagida:1979as,Mohapatra:1979ia}. 
With the deepening of research, our understanding of the Leptogenesis has become increasingly profound, and many aspects have been deeply excavated, such as the thermal effect~\cite{Giudice:2003jh,Anisimov:2010aq}, the flavor effect~\cite{Dev:2017trv},  the wash-out effect~\cite{Garbrecht:2018mrp} and the spectator effect~\cite{Garbrecht:2019zaa}.
Various new insights about the Leptogenesis have been proposed, including Wash-in Leptogenesis~\cite{Domcke:2020quw,Marshak:1979fm}, Lepton-flavorgenesis~\cite{Mukaida:2021sgv}, Afleck-Dine Leptogenesis~\cite{Barrie:2021mwi}, etc. In addition, possible signatures for the Leptogenesis mechanism~\cite{Cui:2021iie,Dror:2019syi} are also explored.

Traditional Leptogenesis mechanism always require explicit ($\mathbf{B-L}$)-violating process(es). In this paper, we propose a new Leptogenesis scenario triggered by the  primordial electron asymmetry in the early universe without requiring any explicit $\mathbf{B-L}$ violation in the theory.  The primordial electron asymmetry can be generated from the decay of a super-heavy Higgs doublet, from the axion-inflation, or from the evaporation of the primordial black hole. 
The key point for this mechanism to work is that the electroweak sphaleron process quenches before the electron Yukawa interaction enters thermal equilibrium at a temperature about $T_e\sim 8.7 \times 10^{4}$ GeV~\cite{Bodeker:2019ajh}. 
In the SM, the electroweak symmetry spontaneous breaking (EWSB) takes place at about $T_{C}\sim159.5\pm1.5$ GeV and the sphaleron quenches at about $T_{\rm sph}^{} \sim 131.7\pm 4.3$ GeV~\cite{DOnofrio:2014rug}. 
However, various studies have shown that the electroweak symmetry may not be restored in the high temperature whenever there is extra Higgs interactions~\cite{Weinberg:1974hy,Mohapatra:1979qt,Fujimoto:1984hr,Dvali:1995cj,Salomonson:1984rh,Bimonte:1995sc,Dvali:1996zr,Orloff:1996yn,Gavela:1998ux,Ahriche:2010kh,Espinosa:2004pn,Bajc:1998jr,Agrawal:2021alq,Meade:2018saz,Baldes:2018nel,Glioti:2018roy,Matsedonskyi:2020mlz,Matsedonskyi:2020kuy,Carena:2021onl,Biekotter:2021ysx,Bai:2021hfb,Matsedonskyi:2021hti,Chao:2021xqv}, leading to the high-scale electroweak symmetry restoration.  In this case, even though the primordial  $\mathbf{B-L}$ asymmetry is totally zero and conserved, the left-handed electron asymmetry can be transported into the B-L asymmetry via the electroweak sphaleron process, while the right-handed electron  asymmetry keeps unchanged until to the $T_e$, after which chiral electron asymmetries is neutralized by the electron Yukawa interaction but the previously produced $\mathbf{B}$ asymmetry keeps un-changed,  since the electroweak sphaleron is quenched, resulting in a non-zero BAU. Numerical studies show that the observed BAU can be addressed by this mechanism. 

The remaining of the paper is organized as follows: in the section II we discuss the electroweak symmetry non-restoration and derive the necessary condition of $T_{\rm sph} \ge T_e$.  We show two workable scenarios in which the BAU is produced by the 
out of equilibrium decay of the Higgs doublet, or by the axion-inflation. The last part is concluding remarks.

\prlsection{Electroweak symmetry non-restoration}{.} \label{sec:nonrestoration} The Higgs vacuum expectation value (VEV) vanishes at a temperature above the 160 GeV due to thermal corrections to the Higgs potential, arising from the Higgs interactions in the plasma with the top quark, the electroweak gauge bosons and the Higgs boson itself,  which means that the electroweak symmetry gets restored at the high temperature in early universe. However, it has been shown by S. Weinberg in the Ref.~\cite{Weinberg:1974hy} that the gauge or global symmetries may not be restored at high temperature and a symmetry non-restoration (SNR) pattern~\cite{Meade:2018saz,Baldes:2018nel,Glioti:2018roy,Matsedonskyi:2020mlz,Matsedonskyi:2020kuy,Carena:2021onl,Biekotter:2021ysx,Bai:2021hfb,Matsedonskyi:2021hti,Chao:2021xqv} could be realized.  

In the minimal SM, the EWSB at the zero temperature is triggered by the negative mass term in the Higgs potential 
\bea
V= -\frac{\mu^2}{2} h^2 +\frac{\lambda}{4} h^4
\eea
where $h$ is the average value of the Higgs field with $\langle h \rangle \approx 246$ GeV and $m_h^{}=126$ GeV at the minimum, $\mu \approx 90$ GeV and $\lambda \approx 0.13$~\cite{ATLAS:2015yey}.  In the early universe there is thermal corrections to the Higgs potential induced by the Higgs interactions with the high-temperature plasma, and the thermal correction to the Higgs mass in the high temperature limit is $\Pi_h \approx 0.4 T^2$~\cite{Quiros:1999jp}.  To realize the  SNR, one only need to extend the SM with new scalar singlet $s$ that couple to the SM Higgs via  $\frac{1}{2}\lambda_{hs}s^2 h^2$~\cite{Meade:2018saz} with a negative quartic coupling, or fermion singlets $\chi$ that couple to the SM Higgs via dimension-5 effective operator $h^2 \bar \chi \chi /\Lambda $~\cite{Matsedonskyi:2021hti}. We focus on the scalar extension case, transforming in a vector representation of the global $O(N_S)$ symmetry~\cite{Meade:2018saz},  then the thermal mass of the SM Higgs is
\bea
\Pi_h = T^2 \left( \frac{\lambda_t^2}{4} + \frac{3g^2}{16} + \frac{g^{\prime 2} }{16} + \frac{
\lambda}{2} + N_s \frac{\lambda_{hs}}{12} \right)
\eea
where $N_s$ is the degree of the scalar singlet.  Taking into account the bounded-from-below constraint for the Higgs potential, one has $|\lambda_{hs}| >\sqrt{\lambda \lambda_s}$, where $\lambda_s$ is the quartic self-coupling of $s$.

There are many cosmological effects induced by the SNR, of which the impact to the electroweak sphaleron may lead to interesting phenomena for the Baryogenesis.   It is well-known that $\mathbf{B}$  changes via the chiral anomaly when the quantum tunneling process occurs between the topologically different degenerate vacua.  In the symmetric phase, the energy barrier for the sphaleron is absent and the sphaleron rate is evaluated as $\Gamma_{\rm sph} \sim \kappa  \alpha_W^5 T^4$~\cite{Arnold:1996dy,DOnofrio:2012yxq}, where $\alpha_W =g^2/(4\pi)$ and $\kappa \sim 100$. In the broken phase, the energy barrier between the topological degenerate vacua grow as VEV increases, and the sphaleron rate is estimated as~\cite{DOnofrio:2012phz} 
\bea
\Gamma_{\rm sph}^{\rm brok} (T) = \kappa_{\rm brok} \alpha_W^4 T^4 \exp\left( - {E_{\rm sph} \over T} \right)
\eea
where $E_{\rm sph}$ is the sphaleron energy proportional to the  Higgs VEV and $\kappa_{\rm brok}$ is a numerical coefficient~\cite{Klinkhamer:1984di}.  Using this formula, one can estimate the temperature that the electroweak sphaleron decouples from the thermal equilibrium by comparing the sphaleron rate $\Gamma/T^3$ with the Hubble rate $H\sim 3T^2/M_P$, where $M_P$ is the Planck mass, which has 
\bea
E_{\rm sph} >  T \log \left(\frac{\kappa_{\rm brok} \alpha_W^4  M_P}{3T} \right) \; .\label{20241118}
\eea
The  sphaleron Energy is first derived by Klinkhamer and Manton~\cite{Klinkhamer:1984di}. In terms of two profile functions $f(\xi)$ and $h(\xi)$, which describe the relevant physical degrees of freedom of the SU(2) gauge fields and the Higgs doublet field, the sphaleron energy is given by 
\bea
E_{\rm sph} &=&\frac{4\pi v} {g} \int_0^\infty d \xi \left[ 4 (f^\prime )^2 + \frac{8}{\xi^2} f^2 (1-f)^2 + \frac{\xi^2}{2} (h^\prime)^2  \right.  \nonumber \\
&&\left.+ h^2 (1-f)^2 +\frac{\xi^2}{16} \sigma^2 (h^2 -1)^2 \right]
\eea
where $v$ is the temperature dependent Higgs VEV,  $\sigma =\sqrt{2\lambda /g^2}$ with $\lambda$ the Higgs quartic coupling, the prime denotes the derivative with respect to $\xi$.  Using the ansatz of $f(\xi)$ and $h(\xi)$ given in the Ref.~\cite{Klinkhamer:1984di}, and further considering that 
\bea
v^2(T) \sim  \frac{\mu^2 -\Pi_h  }{\lambda} \; , 
\eea
one can derive the condition that the Eq.~(\ref{20241118}) holds at a temperature of  ${\cal O}(10^{5})$ GeV,  below which the electron Yukawa interaction enters thermal equilibrium, as
\bea
\lambda_{hs} <- \frac{4.82}{N_s} \; , \label{20241118-2}
\eea
which satisfies all constraints, given a large $N_s$. 

Eq.~(\ref{20241118-2}) provides the necessary condition that the electron Yukawa interaction never reaches thermal equilibrium before the quench of the electroweak sphaleron. However, the electroweak symmetry still need to restored at higher temperature such that the electroweak sphaleron is still an active process that violates $\mathbf{B}$ as required by the first Sakharkov condition.  To get this, one needs to introduce another  scalar singlet $S$, transforming in a vector representation of the global $O(N_s)$ symmetry, with positive quartic coupling to the SM Higgs doublet.  Such that the thermal mass of the SM Higgs turns to be positive at a temperature above the decoupling of $S$ and electroweak symmetry gets restored.  The schemas of this setup can be visually represented as
\begin{center}
\begin{tikzpicture}
\draw[-, thick] (1,-1)--(1.2,-1);
\draw[-, thick] (1,-0.5)--(1.2,-0.5);
\draw[-, thick] (1,1)--(1.2,1);
\draw[-, thick] (1,1.5)--(1.2,1.5);
\node[red, thick] at (2.9,-1) {$T_{\rm EW}^{} $ (traditional)};
\node[red, thick] at (1.8,-0.5) {$m_s$ };
\node[draw,align=center, text width=2.9cm] at (-0.8,-0.5) { $s$: Negative \\ thermal~correction};
\node[red, thick] at (2.6,1) {$T\sim 10^{5} ~{\rm GeV}$ };
\node[red, thick] at (1.8,1.5) {$m_S$ };
\node[draw,align=center, text width=2.9 cm] at (3.8,1.7) {$S$:  Positive \\ thermal~correction};
\node[red, thick] at (1.6,1.9) {$T$ };
\draw[->, thick] (1,-2)--(1,2);
\end{tikzpicture} 
\end{center}
where new particles and their masses are marked in the temperature axis. In this way, the electroweak symmetry is restored in the early universe whenever $S$ is in the thermal bath. As the temperature drops lower,  $S$ is decoupled roughly at $T < m_S/20$,  then the SM Higgs gets a negative thermal mass, leading to the EWSB,  and the electroweak sphaleron is quenched. 

\prlsection{BAU}{.} \label{sec:bau}  In the traditional Leptogenesis mechanism, explicit lepton-number-violating (LNV) process in addition to the electroweak sphaleron is always needed.  These LNV processes  either serve as the source for the generation of the net lepton number, or serve as the spectator process.  
According to the previous discussion, one can see that extra LNV may not be needed in the high-scale electroweak symmetry restoration case,  as the right-handed electron does not  affect transport equations anymore and a total zero initial number density may leads to a net BAU and a electron asymmetry. 
In the following, we will present two workable BAU models: (A) The primordial electron asymmetry be generated from  the axion inflation; and (B) The electron asymmetry be generated from the CP asymmetric decay of heavy Higgs doublets. 

\noindent \framebox[\width]{\textbf{\color{red}~(A):~}} We work in the framework of the SM extended by a gauged $U(1)_{\mathbf{R}}^{}$~\cite{Langacker:2008yv,Chao:2017rwv}, in which the axion inflaton couples to the Chern-Simons term of the new gauge field, i.e., $aF^\prime\tilde{F}^\prime /f_a $. Then the magnetic helicity of the $U(1)_{\mathbf{R}}$ gauge field can be generated during the inflation, which might be transported into non-zero charge densities of various right-handed fermions via triangle anomaly with~\cite{Chao:2024fip}
\bea
\partial_\mu j^\mu_{U} &=& \frac{1}{16\pi^2} \left( -{4\over 9} g^{\prime 2} F \widetilde{F} - g_{\mathbf{R}}^2 F^\prime \widetilde{F}^\prime \right)  \\
\partial_\mu j^\mu_{D} &=& \frac{1}{16\pi^2} \left( -\frac{1}{9}g^{\prime 2} F \widetilde{F} - g_{\mathbf{R}}^2 F^\prime \widetilde{F}^\prime \right)  \\
\partial_\mu j^\mu_{E} &=& \frac{1}{16\pi^2} \left( -g^{\prime 2} F \widetilde{F} - g_{\mathbf{R}}^2 F^\prime \widetilde{F}^\prime \right)  \\
\partial_\mu j^\mu_{N} &=& \frac{1}{16\pi^2} \left( - g_{\mathbf{R}}^2 F^\prime \widetilde{F}^\prime \right) 
\eea
where $g^\prime$ and $g_{\mathbf{R}}^{}$ are the gauge couplings of the $U(1)_Y$ and $U(1)_{\mathbf{R}}$, respectively. Obviously, one has $\partial_\mu (j^\mu_L + j_E^\mu + j^\mu_N -j^\mu_Q-j^\mu_U -j^\mu_D )=0$.  
It has been shown that no BAU can be generated in this case, whenever right-handed neutrinos are Majorana particles, that is similar to the  case of the $U(1)_{B-L}$~\cite{Fukuda:2024pkh}.   While in the case of the high-scale electroweak symmetry restoration, right-handed electron never reaches thermal equilibrium before the quench of the electroweak sphaleron, which means that the non-zero initial number density of the right-handed electron, $n_e^i$, is irrelevant to the BAU, and one only need to consider evolutions of other species.  Considering that the initial number densities are totally zero, one has 
\bea
Y_{B-L_{\slashed{e}}}^{} = + \frac{n_e^i}{s^i} \; , \label{20241126-5}
\eea
where $Y_{B-{L_{\slashed{e}}}}^{}$ is the number density of B-L per comoving volume except that of the right-handed electron,   $s^i$ is the entropy density at the time of the decay of the new gauge field, $n_e^i$ can be calculated following the strategy of Refs.~\cite{Maleknejad:2016qjz,Maleknejad:2020pec}.   The $B-L_{\slashed{e}}$ density derived from the  Eq.(\ref{20241126-5}) is finally converted to the BAU via the electroweak sphaleron process. Below EWPT,  the electron Yukawa interaction enters thermal equilibrium, resetting number densities of chiral electrons  but the BAU will not be changed.  Obviously, the observed BAU can be generated by selecting proper inflation parameters.

\noindent \framebox[\width]{\textbf{\color{red}~(B):~}}  In this scenario, we consider the case where the initial chiral electron asymmetries is generated from the CP asymmetric decay of heavy scalar doublets. New Yukawa interactions can be written as 
\bea
{\cal L} \supset \overline{\ell_{L}^{f}} Y_{fg}^i \Phi_i E_{R}^g + h.c. \label{20241125-1}
\eea
where $\Phi_i$ ($i=1,2$) is heavy Higgs doublets, $\ell_L^{f}$ is the left-handed lepton doublet  of flavor $f$, $E_R^g$ is right-handed charged lepton of flavor $g$,  $Y_{fg}^{i}$ are Yukawa coupling matrices. 

We consider the case where the reheating temperature of the universe is higher than $M_{\Phi_i}$ such that $\Phi_i$ is thermally populated initially. The CP-violating (CPV) decay of the heavy Higgs doublet will generate a charge asymmetry, which is similar to the case of the thermal Leptogenesis and the GUT Baryogenesis. The mass hierarchy for $\Phi_i$ is taken as $M_{\Phi_1} <M_{\Phi_2}$, such that the charge asymmetry generated by $\Phi_2$ will be washed-out by the inverse decay process $E_R^g \overline{\ell_L^f} \to \Phi_1$ and we only need to consider the CPV decay of $\Phi_1$.   The CP asymmetry, $\varepsilon$, is generated from the interference between the tree-level and the one-loop level (self-energy ) Feynman diagrams~\cite{Mukaida:2024eqi}.  Here we take it as the free parameter since it is a function of $Y_{fg}^i$ and $M_{\Phi_i}$ that are totally unconstrained parameters.

\begin{figure*}[t]
        \includegraphics[width=8.5cm]{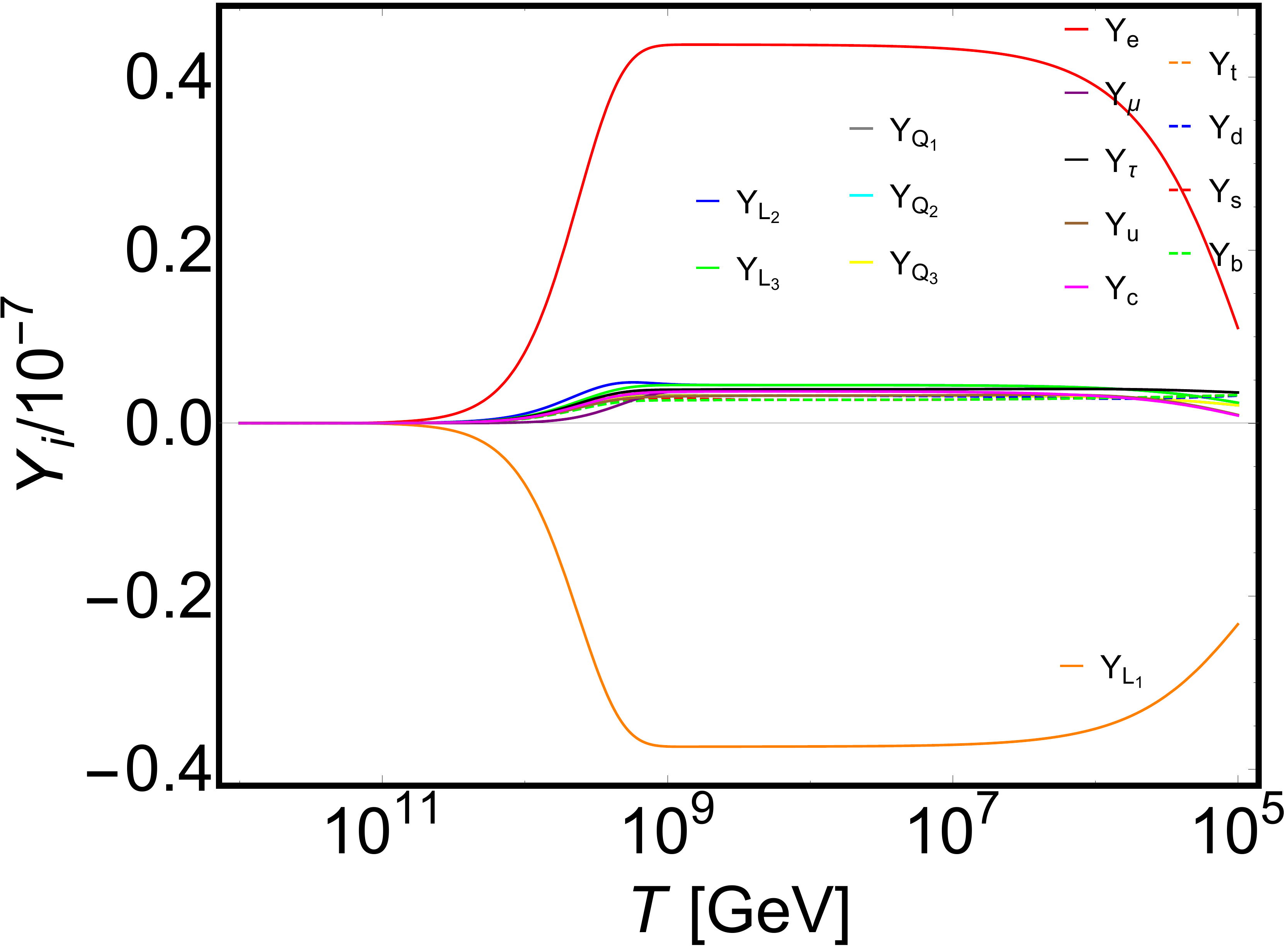}
	\includegraphics[width=8.5cm]{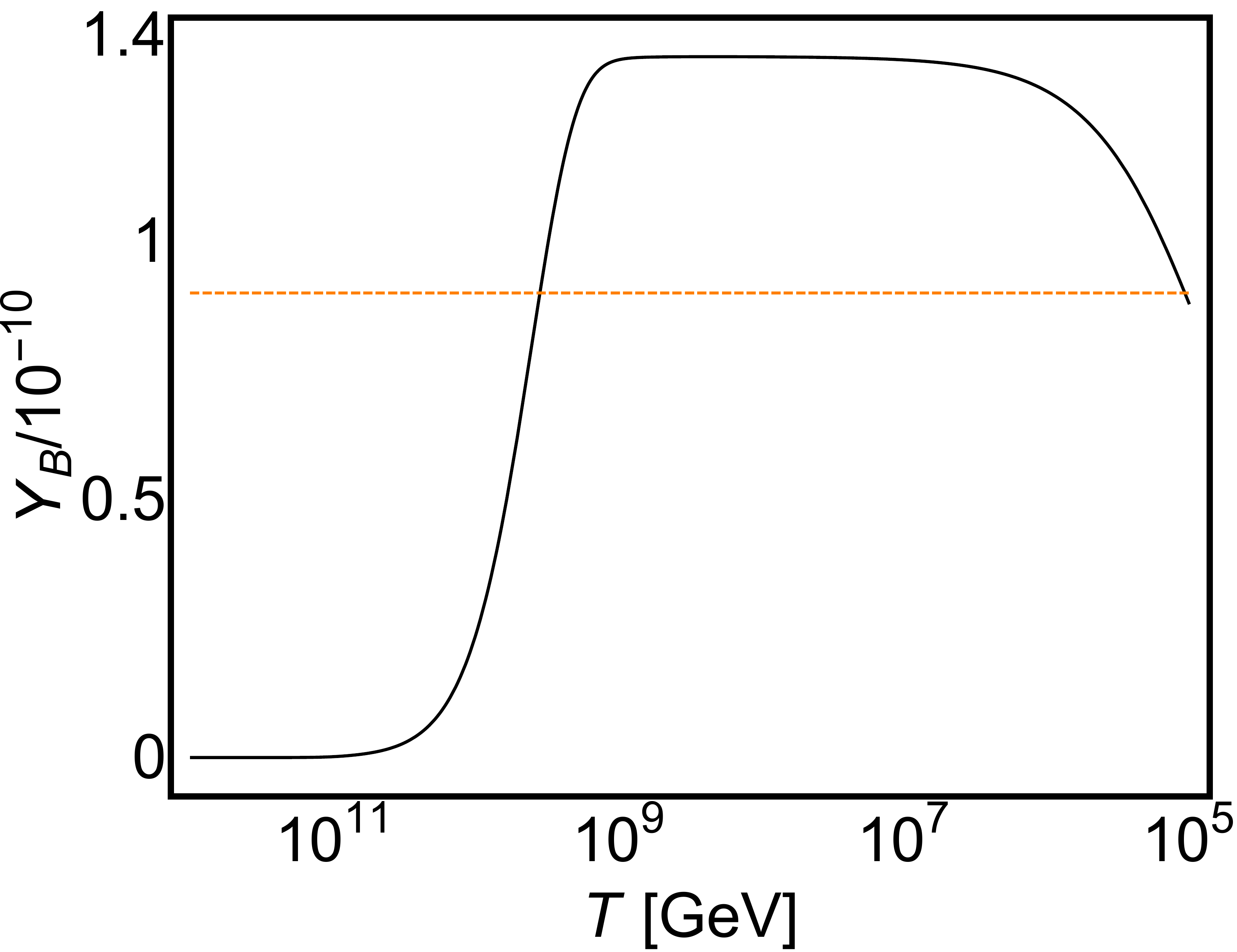}
	\caption{ Left-panel: $Y_i =\eta_i/T$ as the function of $T$, by setting $m_\Phi =10^{10}$ GeV  $Y\sim 0.05$ and $\varepsilon=1\times 10^{-6}$. Right-panel $Y_B$ as the function of the $T$ with inputs the same as those in the plot of the left-panel. The horizontal dashed line is the observed BAU. }
	\label{etaB}
\end{figure*}

To write down transport equations, we define $Y_\Sigma = (n_\Phi^{}  + n_{\bar \Phi}^{} )/s$ with $s$ the entropy density of the universe,  and further take $\eta_\Phi$ and $\eta_{L_k}^{}$ as chemical potentials of $\Phi$ and $L_k$, respectively.
Relevant Boltzmann equations for the heavy Higgs doublet are
\bea
-\frac{d Y_\Sigma^{} }{d \ln T} &\approx& -\frac{\gamma_D}{H}  \left( Y_\Sigma^{} - Y_\Sigma^{\rm EQ}\right)  \label{20241121-0}\\
%
%
-\frac{d}{d \ln T} \left(  \mu_\Phi \over T \right) &= &-2 \frac{\gamma_D}{H}  \left( \frac{\mu_\Phi} {T} -\frac{4}{3}\frac{\mu_{L_k}}{T} \right)  \label{20241121-1}
\eea
where we have neglected the effect of two-Higgs annihilation in the Eq.~(\ref{20241121-0}) for simplicity,  and 
\bea
\gamma_D = {K_1^{}(z) \over K_2^{} (z)}  \Gamma_\Phi^{} 
\eea
with $K_i(z)$ the Bessel-K function of rank ``i" and $\Gamma_\Phi$ the decay rate of $\Phi$. The transport equation for the lepton doublet $L_k$ that participates the Yukawa interaction in Eq.~(\ref{20241125-1}) is
\bea
-\frac{d}{d\ln T}\left( \frac{\mu_{L_k}}{T}\right) &=&-\frac{1}{g_{L_k}^{} }\frac{\gamma_{\rm WS}^{}}{H}\left[\sum_{i=1}^3 \left(  \frac{\mu_{L_{i}}}{T}+3 \frac{\mu_{Q_{i}}}{T} \right) \right]\nonumber\\
&&-\frac{1}{g_{L_k}}\frac{\gamma_{Y_{E_k}}}{H}\left(-\frac{\mu_{E_k}}{T}+\frac{\mu_{L_k}}{T}-\frac{\mu_H}{T}\right) \nonumber \\
&& + {1 \over g_{L_k}^{} }\frac{4\pi^2 g_{*s}^{}}{15}  \frac{\gamma_D}{H}  \varepsilon  \left( Y_\Sigma^{}  - Y_\Sigma^{\rm EQ}  \right) \nonumber \\
&&-\frac{2 }{g_{L_k}} g_\Phi^{} \frac{\gamma_D}{H}  \left(  \frac{4}{3} \frac{\mu_{L_k}}{T}-   \frac{\mu_\Phi }{T} \right)   \label{20241129-1}
\eea
where $\gamma_{\rm WS}^{} $ and $\gamma_{Y_{E_k^{}}}$ are interaction rates of the the electroweak sphaleron and  the Yukawa interaction, respectively. Transport equations for other species are the same as those in the SM.

Combing Eqs~(\ref{20241121-0}), (\ref{20241121-1}), (\ref{20241129-1}) with transport equations of other species given in the Ref.~\cite{Domcke:2020kcp}, one can estimate the BAU induced by this model. The final BAU is related to chemical potentials of the baryon number as 
\bea
Y_B=  \frac{n_B}{s} =\frac{15}{4\pi^2 g_{*s}^{}(T_0) } \frac{\mu_B}{T}
\eea
where $T_0$ is the temperature of EWPT.  We show in the left-panel of the Fig.~\ref{etaB} chemical potentials of various SM particles as the function of the temperature by setting $m_\Phi =10^{10}$ GeV, ${\cal O}(Y)\sim 0.05$ and $\varepsilon=1\times 10^{-6}$. For the physical meaning of each curve, please pay attention to the annotations in the figure. One can see that nonzero baryon chemical potentials can be generated in this case even if the initial B-L is totally zero.  We show in the right-panel of the Fig.~\ref{etaB}, $Y_B$ as the function of the temperature  with  inputs  the  same as those in the plot of the left-panel. The horizontal line is the observed BAU. One can immediately conclude that the BAU can be addressed in this case by select a proper CPV source term.  
For the shape of the curve, non-zero $\eta_L$ and $\eta_e$ are generated during the decay of the heavy Higgs doublet, where $\eta_L$ is immediately transport into nonzero $\eta_B$ via the electroweak sphaleron process that is in the thermal equilibrium.    
As the temperature drops below $10^{7}$ GeV, the electron Yukawa interaction starts to affect the evolution of the of $Y_B$, even though it never reach thermal equilibrium before the EWPT. 
It leads to the decrease of the $Y_B$, but still large enough.   As the temperature drops below the $T_{\rm EW}^{}$, the electroweak sphaleron is quenched and the  generated BAU will not be changed. When the electron Yukawa interaction enters thermal equilibrium, chiral electron asymmetries will be neutralized with each other resulting a net electron asymmetry.

\prlsection{Summary}{.} The BAU is a longstanding problem in the high-energy physics and cosmology. Many workable Baryogenesis models were proposed based on three Sakharov conditions, most of which require explicit (B or L)-violating-process.  For example, the traditional Leptogenesis mechanism requires the L-violating-process mediated by the heavy seesaw particles. The Afeck-Dine Baryogenesis  mechanism requires explicit  B (or L)-violating term in the potential. In this letter, we point out that a successful Leptogenesis in a electroweak symmetry non-restoration scenario may neither need extra explicit B(or L)-violating process nor need extra spectator process beyond the electroweak sphaleron.  The reason is that the right-handed electron is roughly decoupled from the transport equations.  We have showed two workable scenarios: (1) The axion inflation Baryogenesis in the $U(1)_R$ model, and (2) Leptogenesis induced by CPV decay of  heavy scalar doublets into electrons. It should be mentioned that this mechanism may work in many models,  such as the axion-inflation Baryogeneis with the $U(1)_L$ or the $U(1)_{B-L}$ extensions,  or any model that contains CPV process related to right-handed electrons. In summary, this Leptogenesis mechanism may evade the constraints imposed by seesaw particles, potentially accommodating a broader range of model constructions.

\label{sec:summ}

\begin{acknowledgments}

This work was supported in part by the National Key R\&D Program of China under Grant No. 2023YFA1607104, by the National Natural Science Foundation of China under Grants No. 11775025 and No. 12175027,  and by the Fundamental Research Funds for the Central Universities under Grant No. 2017NT17.

\end{acknowledgments}

\bibliography{references}

\end{document}